# High-resolution long-range 3D single-photon imaging with a compact SPAD array

Zunwang Bo[1,2]*, Chenjin Deng[1,2], Fei Wang[2], Wenlin Gong[3], Yuanhao Su[4], Yichen Zhang[2], Mingliang Chen[1,2], Chunfang Wang[4], Shensheng Han[1,2]

*1. Key Laboratory for Quantum Optics, Shanghai Institute of Optics and Fine Mechanics, Chinese Academy of Sciences，Shanghai 201800, China*

*2. Wangzhijiang Innovation Center for Laser, Aerospace Laser Technology and System Department, Shanghai Institute of Optics and Fine Mechanics, Chinese Academy of Sciences, Shanghai 201800, China*

*3. School of Optoelectronic Science and Engineering, Soochow University, Suzhou 215006, China*

*4. College of Science, Physics Department,University of Shanghai for Science and Technology, Shanghai 200093, China*

*Corresponding Author: zwbo@siom.ac.cn*

**Abstract:**

High-resolution three-dimensional imaging under photon-starved conditions remains challenging. Here, we demonstrate a high-resolution long-range 3D single-photon imaging system based on a digital micromirror device (DMD) and a compact 64 × 64 single-photon avalanche diode (SPAD) array. By combining high-resolution spatial modulation with parallel time-resolved detection, the system extends the effective spatial sampling beyond the native detector format while preserving depth information through time-of-flight measurement. In outdoor experiments at a stand-off distance of 670 m, we achieved 3D reconstruction of natural targets with an effective spatial resolution of 256×256. These results validate the proposed method as an effective approach for high-resolution long-range 3D single-photon imaging using compact SPAD arrays.

## 1. Introduction

Three-dimensional imaging under photon-starved conditions is of broad interest for applications such as remote sensing, topographic mapping, and autonomous driving.

[1-4]. In such scenarios, the returning signal is often extremely weak because of large propagation loss and limited target reflectivity, making high-quality imaging highly challenging. Single-photon imaging is particularly attractive for these applications because it combines extreme sensitivity with time-resolved depth sensing[5-10]. Among existing approaches, scanning single-photon lidar has demonstrated impressive long-range performance[8], but its reliance on sequential scanning can limit acquisition efficiency and increase system complexity. Array-based imaging with single-photon avalanche diode (SPAD) sensors provides an alternative route by replacing point-by-point scanning with parallel detection[9,10]. However, the spatial resolution of current SPAD arrays remains constrained by their native pixel count, because per-pixel integration of high-precision time-to-digital converter (TDC) circuits places a substantial burden on data throughput and power consumption, which makes large-scale time-resolved pixel integration highly challenging[11]. As a result, there remains a fundamental challenge in achieving both high spatial resolution and efficient three-dimensional acquisition with compact single-photon detector arrays.

Computational imaging offers an alternative by removing the direct one-to-one correspondence between detector pixels and image pixels[12]. Among these approaches, ghost-imaging and single-pixel-based architectures are especially appealing for photon-limited sensing because they encodes high-dimensional spatial information into low-dimensional photon-counting measurements for computational recovery, in principle even with a single-pixel bucket detector[13-15]. In practice, however, successful reconstruction requires accurate measurement of the intensity fluctuations of the returned photons. Under extremely weak-light conditions, the return signal typically has a very low signal-to-noise ratio, which severely degrades reconstruction quality. Although repeated measurements and signal accumulation can improve image fidelity, they usually require prohibitively large sample numbers and thus lead to very low imaging efficiency. Reconstruction methods based on learned priors have also been explored for low-sampling ghost imaging, but most demonstrations remain limited to simple, highly structured targets in controlled indoor environments. As a result, conventional ghost-imaging schemes have not yet fully realized their advantage of recovering high-resolution images from low-dimensional detection for photon-starved imaging of complex natural scenes.

To overcome this limitation, we present a high-resolution long-range 3D single-photon imaging architecture that combines high-resolution spatial modulation with compact time-resolved array detection. Related strategies have previously been explored in infrared compressive imaging and continuous-light single-photon imaging[15-17]. By using a digital micromirror device for blockwise spatial encoding and a compact SPAD array for parallel photon detection, the proposed method extends the effective spatial sampling beyond the native detector format while preserving depth information through time-of-flight measurement. In outdoor experiments at a stand-off distance of 670 m, we demonstrate 3D reconstruction of natural targets with an effective spatial resolution of 256×256. Compared with direct imaging using the native 64×64 detector format under the same measurement conditions, the proposed method provides substantially improved spatial detail and image contrast. These results establish a practical route toward high-resolution long-range 3D single-photon imaging using compact SPAD arrays.

**2. Principle and system implementation**

Figure 1 illustrates the operating principle of the proposed high-resolution long-range 3D single-photon imaging system. A pulsed laser beam is first expanded and shaped to provide approximately uniform illumination over the target scene. The reflected photons are collected by a receiving telescope, which forms an optical image of the target on a high-resolution digital micromirror device (DMD). A SPAD array is positioned behind the DMD, and the DMD modulation plane is optically conjugated to the detector plane so that a one-to-one spatial correspondence is established between the two planes. This configuration combines high-resolution spatial encoding with parallel time-resolved photon detection.

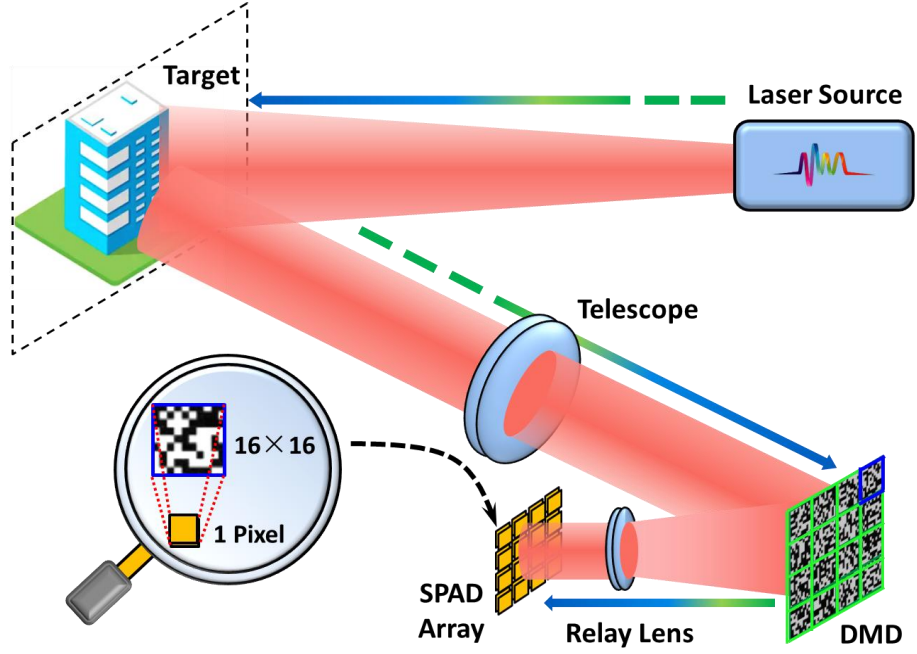

Figure 1 Schematic of the proposed high-resolution 3D single-photon imaging system. A pulsed laser illuminates the target scene, and the reflected light is collected by a receiving telescope and imaged onto a digital micromirror device (DMD). The DMD performs high-resolution spatial modulation of the target image, and each 16×16 DMD block is mapped onto one pixel of the compact SPAD array for parallel photon detection

The key idea is to use each SPAD pixel as an independent coded measurement channel for a local image patch. The image formed on the DMD is divided into non-overlapping subregions, each mapped onto one SPAD pixel. In the present implementation, one SPAD pixel corresponds to a $16 \times 16$ block of DMD micromirrors. During acquisition, the DMD sequentially loads a set of spatial patterns, and each SPAD pixel records the photon counts associated with the modulation of its corresponding local block. Since all SPAD pixels operate simultaneously, the system performs parallel blockwise acquisition over the entire field of view. For the $k$-$th$ SPAD pixel, the measurement process can be expressed as:

$$y_k = A_k x_k + n_k \qquad (1)$$

Where $x_k$ denotes the vectorized reflectivity of the corresponding local scene patch, $A_k$ is the sensing matrix determined by the loaded DMD patterns, $y_k$ is the photon-count measurement vector, and $n_k$ accounts for photon noise, dark counts, and background noise. Each detector pixel therefore provides a coded low-dimensional measurement of a local high-resolution patch.

A 1024 × 1024 region of the DMD is used for modulation. To reduce the number of unknowns and improve robustness under photon-starved conditions, each 4 × 4 group of neighboring micromirrors is treated as one effective modulation cell. The effective modulation grid is therefore 256 × 256. Under the 64 × 64 DMD-to-SPAD mapping, each SPAD pixel reconstructs a 4 × 4 local patch, and the reconstructed patches from all detector pixels are stitched to form a final image with an effective spatial resolution of 256×256. In this way, the spatial sampling of the reconstructed image is extended well beyond the native detector format while retaining the parallelism of array detection.

Three-dimensional imaging is enabled by the time-resolved capability of the SPAD array. For each detector pixel, the photon arrivals are recorded as a time-of-flight histogram with discrete temporal bins. The blockwise reconstruction is then applied independently to each temporal slice, yielding a set of depth-resolved images that together form a 3D image cube. The initial 3D image is then refined using inverse 3D deconvolution based on the modified SPIRALTAP solver to suppress noise and block artifacts and to exploit the spatial correlation of natural scenes[7,20]. The final estimate is obtained by solving:

$$\hat{O} = \arg\min_{O}\left\{-\log \Pr(Y;O) + \lambda \|O\|_{TV}\right\} \tag{2}$$

where $Y$ is the 3D image by blockwise reconstruction.

**3. Experimental setup**

The proposed system was experimentally implemented for long-range active single-photon imaging at a stand-off distance of 670 m. A pulsed laser (Connect CoLID-I-1550-B) operating at 1550 nm was used as the illumination source, with a pulse width of 1 ns full width at half maximum (FWHM), a repetition rate of 25 kHz, and a pulse energy of 100 μJ. The outgoing beam was expanded and shaped by a launch lens with a 50 mm aperture and a focal length of 100 mm to provide approximately uniform illumination over the target scene.

The backscattered photons were collected by a receiving telescope with an aperture of 150 mm and a focal length of 675 mm. The collected target image was formed on a digital micromirror device (DMD, Vialux V-9501VIS), whose native resolution is 1920 × 1080 with a micromirror pitch of 10.8 μm. In the present experiment, a 1024 × 1024 region of the DMD was selected for spatial modulation.

The modulated return signal was detected using an InGaAs 64 × 64 SPAD array (CETC GD5551-C), which offers a detection efficiency of 20% at 1550 nm and a dark count rate of 2 kHz. Each pixel integrates a time-to-digit converter (TDC) unit that uses a 12-bit counter, allowing time stamps to be recorded with a time resolution of 1 ns. The DMD plane and the SPAD array plane were optically matched such that each SPAD pixel corresponded to a 16 × 16 block of DMD micromirrors. To reduce the number of independent unknowns in the reconstruction, every 4 × 4 DMD micromirrors were grouped into one effective modulation cell. Therefore, although the selected DMD region contained 1024 × 1024 micromirrors, the effective number of independently modulated cells was 256 × 256.

## 4. Results

The proposed method was first evaluated on a long-range outdoor natural target. Figure 2 shows representative reconstruction results for a television tower at a stand-off distance of 670 m. Owing to the large vertical extent of the target, a single acquisition with a diagonal field of view of 1.3° could capture only part of the tower. To recover the overall three-dimensional structure, five acquisitions were performed sequentially from the bottom to the top of the tower at different vertical pointing positions, and the reconstructed results were stitched to form the composite 3D rendering shown in Fig. 2(b). This stitched image corresponds to most of the main body of the tower shown in Fig. 2(a). In each acquisition, 256 spatial modulation patterns were loaded onto the DMD, and the photon signal for each pattern was accumulated over 240 laser pulses to improve the detection signal-to-noise ratio. Under these conditions, the photon counting rate for pixels containing target echoes was approximately 10%, which means that a photon was detected in about 10 out of every 100 laser shots. The result from a single acquisition is shown in Figs. 3 and 4.

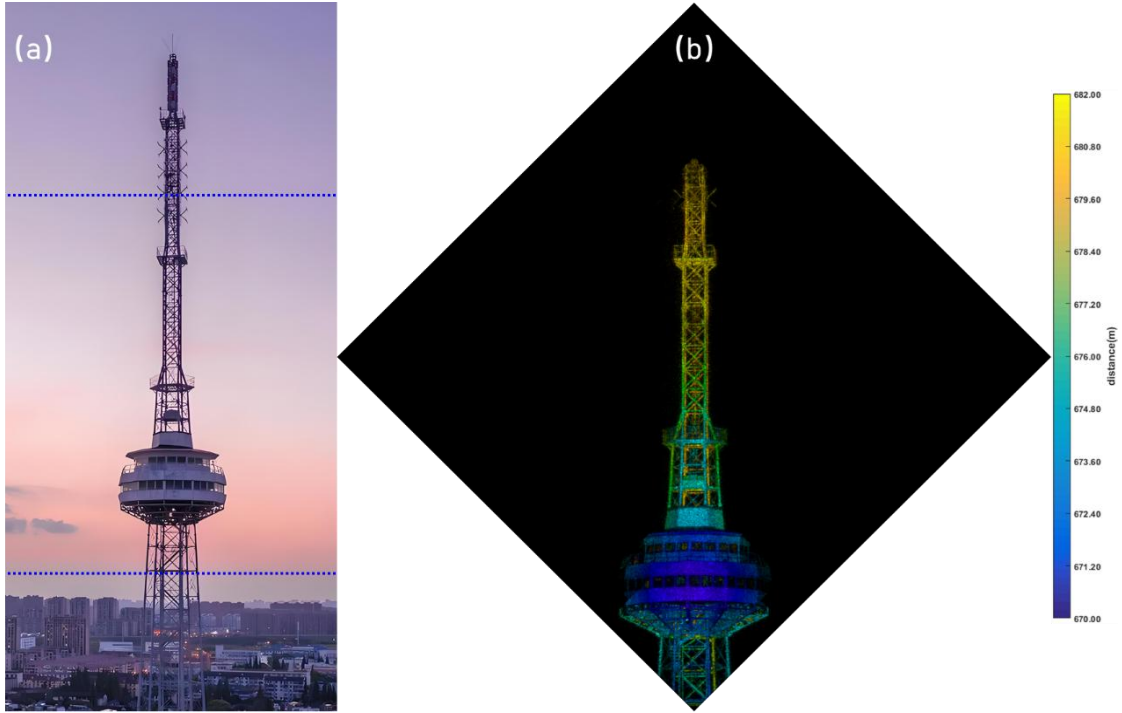

Figure 2 Three-dimensional reconstruction of a television tower at a stand-off distance of 670 m. (a) Photograph of the television tower. (b) Composite 3D rendering of the tower obtained by stitching together five sequential acquisitions performed from bottom to top. The color scale represents depth information.

Figures 3 and 4 show representative results from two different fields of view of the television tower, corresponding to the main body of the tower and the upper steel-frame structure, respectively. In each case, the proposed high-resolution reconstruction is compared with direct focal-plane imaging using the native 64 × 64 SPAD array under identical acquisition conditions. Results are presented for 80, 240, and 480 pulse accumulations per spatial modulation pattern. With 256 spatial modulation patterns, these settings correspond to total acquisition times of 0.82s, 2.46s, and 4.92s, respectively.

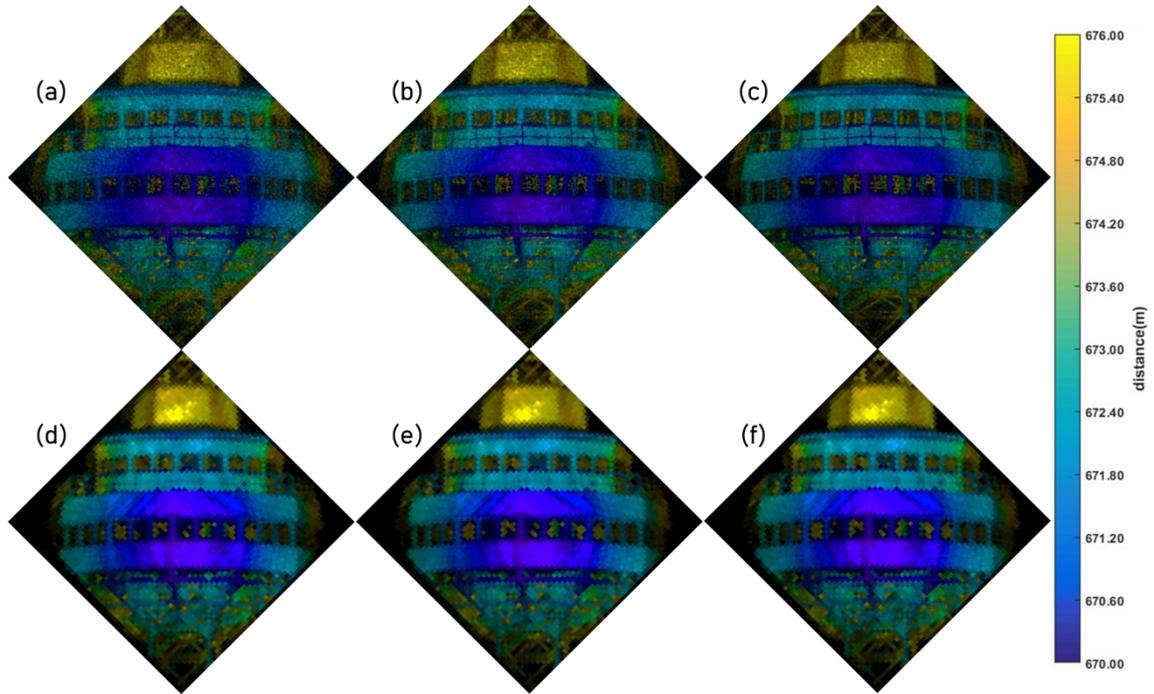

Figure 3 Imaging results for the main body of the television tower (the second acquisition from bottom to top). (a)–(c) High-resolution reconstruction results obtained with 80, 240, and 480 pulse accumulations for each spatial modulation pattern, respectively. (d)–(f) Corresponding direct focal-plane images acquired by the native 64 × 64 SPAD array with the same pulse accumulation numbers. The color scale represents depth information.

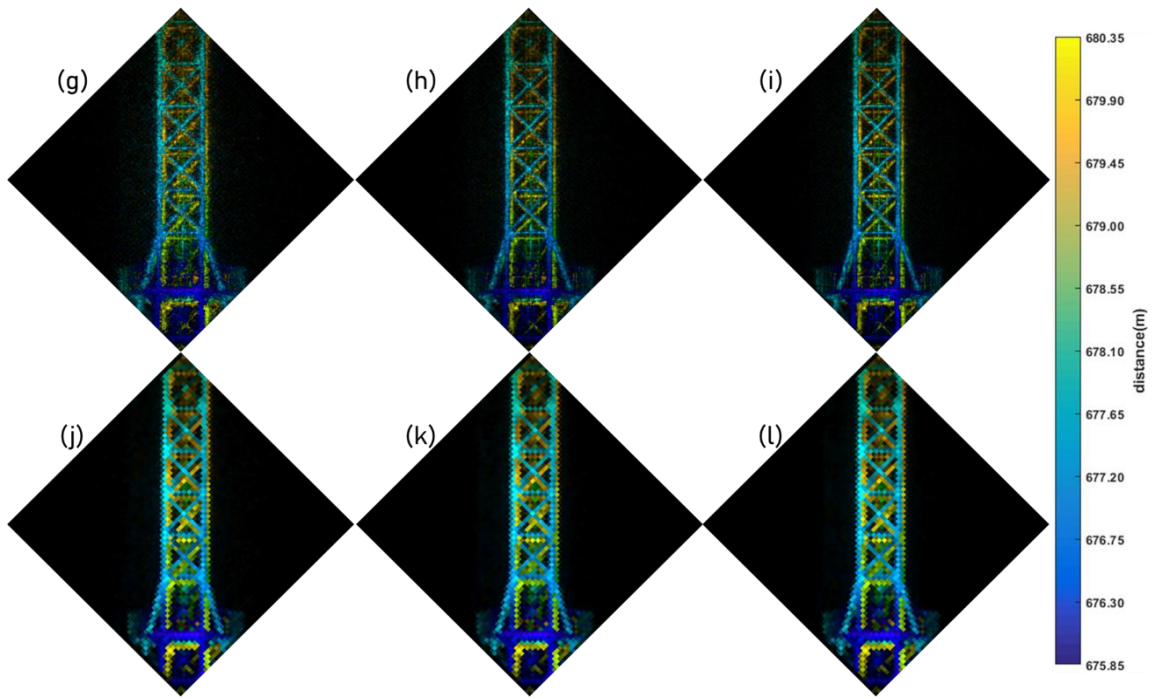

Figure 4 Imaging results for the upper steel-frame structure of the television tower (the third acquisition from bottom to top). (g)–(i) High-resolution reconstruction results obtained with 80,

240, and 480 pulse accumulations for each spatial modulation pattern, respectively. (j)–(l) Corresponding direct focal-plane images acquired by the native 64×64 SPAD array with the same pulse accumulation numbers. The color scale represents depth information.

A comparison of the results in Figures 3 and 4 shows that the proposed method provides a substantial improvement in spatial resolution over direct imaging with the native 64 × 64 SPAD array. In the direct focal-plane images, the target is limited by the coarse detector sampling, and only the overall outline of the tower can be recognized. By contrast, the proposed high-resolution 3D imaging results recover substantially more structural detail. In particular, fine features that are difficult to distinguish in the direct SPAD images, such as handrails, steel tubes, and the three-dimensional steel-frame structure, are clearly resolved in the reconstructed results. These observations demonstrate that the proposed architecture can effectively extend the imaging capability of the compact SPAD array beyond its native detector format.

As the number of accumulated pulses per spatial modulation pattern increases, the signal-to-noise ratio of the reconstructed high-resolution 3D images is further improved. The results obtained with 480 pulse accumulations exhibit the highest image quality among the three cases. However, the results obtained with 240 pulse accumulations, corresponding to a total acquisition time of 2.46 s, already provide a clear and accurate representation of the three-dimensional structure of the television tower. This indicates that the proposed method can achieve a favorable balance between image quality and acquisition time, while maintaining a clear advantage over conventional direct focal-plane imaging.

## 5. Discussion

The effectiveness of the proposed method arises from the parallel and localized nature of the detection process. Each SPAD pixel serves as an independent measurement channel for only a limited local subregion, rather than encoding the entire scene into a single channel. This blockwise parallel detection reduces information aliasing and limits the degree of resolution extension required for each detector channel, thereby improving reconstruction robustness under low signal-to-noise ratio conditions.

In this sense, the proposed architecture provides a practical route for overcoming the pixel-count limitation of compact time-resolved detector arrays, especially for imaging systems in which large-scale per-pixel temporal readout remains constrained by bandwidth, circuit complexity, and power consumption.

The passive imaging experiment under solar illumination further supports the generality of the architecture. Under the same sampling conditions and acquisition time as those used for the active-imaging results in Figs. 2–4, imaging of the Sheraton Hotel building at a distance of 2.0 km again showed clear improvements in spatial resolution and image contrast over direct focal-plane imaging with the native 64 × 64 SPAD array. These results suggest that the proposed approach is not limited to active pulsed 3D imaging, but may also be extended to passive long-range high-resolution imaging. Future work will focus on improving photon efficiency, exploiting global scene correlations more effectively in reconstruction, and extending the method to more challenging passive and dynamic scenarios.

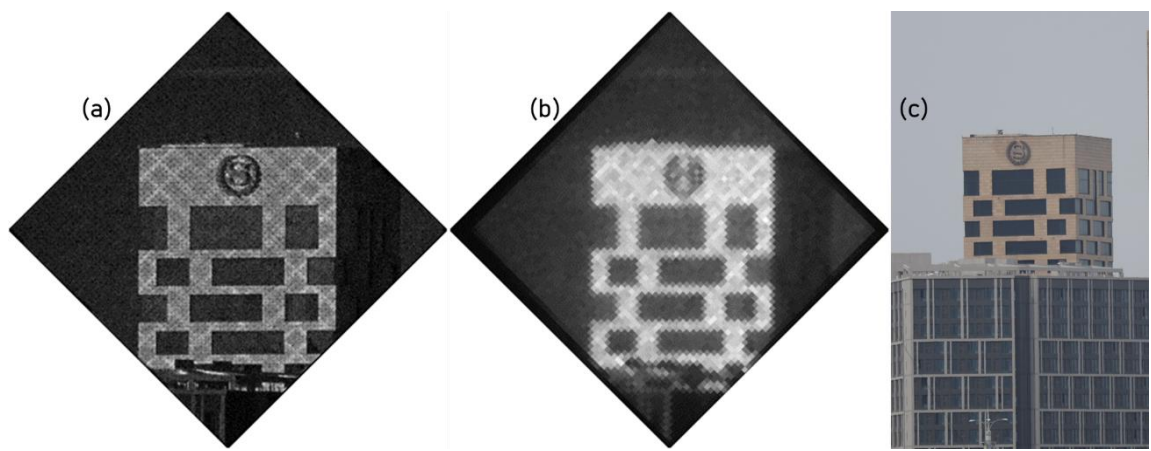

Figure 5 Passive high-resolution single-photon imaging of the Sheraton Hotel building under solar illumination at a stand-off distance of 2.0 km. The sampling conditions and acquisition time were the same as those used for the active-imaging results in Figs. 2–4. (a) Reconstructed image obtained with the proposed method. (b) Direct focal-plane image acquired by the native 64 × 64 SPAD array. (c) Photograph of the target building.

## 6. Conclusion

We have demonstrated a high-resolution long-range 3D single-photon imaging architecture that combines high-resolution spatial modulation with a compact 64×64 SPAD array. By using the detector array as parallel coded measurement channels, the proposed method extends the effective spatial sampling beyond the native detector format while preserving depth information through time-of-flight measurement.

In outdoor experiments at a stand-off distance of 670 m, the method successfully reconstructed the three-dimensional structure of real targets and recovered structural details that were difficult to resolve with direct SPAD focal-plane imaging. Clear 3D imaging was achieved with a total acquisition time of 2.46 s per field of view, indicating a favorable balance between image quality and acquisition speed. These results show that compact time-resolved detector arrays, when combined with optical coding and computational reconstruction, can provide a practical route toward high-resolution long-range single-photon imaging.